\documentclass[journal]{IEEEtran}
\ifCLASSINFOpdf
\else
\fi
\usepackage{cite}
\usepackage[caption=false,font=footnotesize]{subfig}
\usepackage{graphicx}
\usepackage[cmex10]{amsmath}

\usepackage{algorithm}
\usepackage{algpseudocode}
\usepackage{amsmath,amssymb}

\usepackage{stfloats}

\usepackage{color}
\usepackage{placeins}
\usepackage{float}
\usepackage[colorlinks=true,linkcolor=black,citecolor=black,urlcolor=black]{hyperref}
\usepackage{tabularx,colortbl,multirow}
\usepackage{enumerate}
\usepackage{enumitem}
\usepackage{xcolor}

\newcommand{\rev}[1]{#1}

\usepackage{multirow}
\usepackage{makecell}

\usepackage{bm}

\renewcommand\eqref[1]{(\ref{#1})}

\usepackage{tikz}
	\usetikzlibrary{shapes,decorations,decorations.pathreplacing}
	\usetikzlibrary{arrows,shapes,positioning,calc,fadings,patterns}
	\usetikzlibrary{decorations,decorations.pathreplacing,decorations.pathmorphing}
\usepackage{tikz-3dplot}

\usepackage{booktabs}     
\usepackage{threeparttable}
\usepackage{amssymb}

\begin{document}
\title{
Spatio-Sequential Recurrent Network for 3-D Tunnel Propagation  Modeling
}

\author{Jiahao~Li,~Jingxin~Xue,~Keqi Ni, Kunyu~Wu,~Hao~Qin,~\IEEEmembership{Member,~IEEE},~Xinyue~Zhang, and~Xingqi~Zhang,~\IEEEmembership{Senior~Member,~IEEE}
	\thanks{Jiahao Li, Jingxin Xue, Keqi Ni, Kunyu Wu and Hao Qin are with the School of Electronics and Information Engineering, Sichuan University, Chengdu, China.}
    \thanks{Kunyu Wu is also with the Electrical Engineering Division, Department of Engineering, University of Cambridge, Cambridge CB3 0FA, U.K.}
    \thanks{Xinyue Zhang is with the School of Electrical and Electronic Engineering, University College Dublin, Dublin, Ireland.}
    \thanks{Xingqi Zhang is with the Department of Electrical and Computer Engineering, University of Alberta, Canada T6G 2H5.}
}

\markboth{IEEE ANTENNAS AND WIRELESS PROPAGATION LETTERS}%
{LI \textit{et al.} \MakeLowercase{\textit{}}: }


\maketitle

\begin{abstract}
Fine-mesh parabolic wave equation (PWE) simulations are high-fidelity but time-consuming, which limits real-time tunnel propagation analysis and motivates coarse-to-fine reconstruction. Existing machine learning (ML)-assisted tunnel models typically provide only one-dimensional (1-D) longitudinal refinement or two-dimensional (2-D) cross-sectional refinement, rather than joint 3-D enhancement.
 Motivated by this gap, this letter proposes a U-shaped gated spatio-sequential recurrent neural network (UG-SSRNN), a spatio-sequential reconstruction model for tunnel electromagnetic fields. UG-SSRNN jointly super-resolves transverse slices and models longitudinal evolution. It uses sliding-window context encoding and a K-layer convolutional recurrent backbone with a shared propagation-context state and diagonal feedback. A prediction-aware upsampling head leverages the previous prediction to improve slice-to-slice consistency. \rev{Experiments on four tunnel cross sections, unseen-material and unseen-frequency tests, and validation in the Massif Central tunnel show close agreement with fine-mesh PWE references. The proposed approach significantly reduces tunnel electromagnetic modeling time.}

\end{abstract}

\begin{IEEEkeywords}
Tunnel propagation, parabolic wave equation, recurrent neural network, spatio-sequential reconstruction 
\end{IEEEkeywords}

\IEEEpeerreviewmaketitle

\section{Introduction}
In railway intelligent transportation systems (ITS), the performance of communication-based train control (CBTC) systems critically depends on accurate modeling of radio wave propagation in tunnel environments~\cite{9284448, 7836324, 6387578, 6808529}. Many deterministic solvers are derived from the parabolic wave equation (PWE), which have been widely adopted for simulating field evolution in long guiding structures such as railway tunnels~\cite{11177968, 7875498, 8487053, OZGUN20112638}. As operating frequencies increase, finer grid resolution is required in both transverse and longitudinal directions, leading to a substantial rise in the number of marching steps and overall computational cost. This renders fine-mesh PWE solvers impractical for time-critical applications and real-world engineering deployment~\cite{Huang24Frequency, 11198828, 11186207}. Consequently, there is an urgent need for faster yet physically reliable modeling approaches tailored to tunnel propagation analysis.

Machine learning (ML) techniques have been actively explored to reduce computational burden while preserving modeling fidelity~\cite{deng2025marsradiomapsuperresolution, 9713743, 9496115, 10005074}. In tunnel scenarios, one kind existing mainstream methods largely operate in a coarse-to-fine manner at one-dimensional or two-dimensional levels. Specifically, one stream of approaches uses partial/coarse axial information to predict corresponding fine-grid axial profiles \cite{Hao2023rnn, https://doi.org/10.1049/iet-map.2019.0988}. Another stream takes coarse cross-sectional fields as input to reconstruct fine cross-sectional fields as output \cite{Qin24DBP}. 
Despite their efficiency, these methods typically decouple the 3-D propagation physics, either neglecting transverse spatial details or ignoring longitudinal evolutionary consistency.
Furthermore, a straightforward extension to full 3-D convolutional neural networks remains challenging, as volumetric models are computationally expensive and memory intensive, and are difficult to train on long, high-aspect-ratio tunnel volumes~\cite{Ji20133DCNN, Tran20153DCNN}. 

\rev{To address these limitations, we propose a U-shaped gated spatio-sequential recurrent neural network (UG-SSRNN), a 3-D spatio-sequential recurrent framework for tunnel electromagnetic field reconstruction.} The model jointly performs longitudinal propagation modeling and transverse spatial refinement within a unified coarse-to-fine formulation. Along the longitudinal direction, field evolution is formulated as sequential prediction over depth-ordered slices. For each transverse cross section, spatial super-resolution is applied to refine fine-scale details. In addition, a sliding-window strategy fuses neighboring coarse slices to reconstruct each target fine-mesh slice, which improves depth-wise consistency and suppresses numerical artifacts.

The proposed framework is particularly well suited to tunnel propagation modeling, as it explicitly accounts for the sequential nature of wave propagation, the sparse availability of coarse-grid observations, and the need for consistent refinement across longitudinal positions, thereby achieving robust coarse-to-fine reconstruction over long propagation distances. It has been systematically evaluated across four representative tunnel scenarios, demonstrating superior accuracy, generalization capability, and computational efficiency. In addition, engineering validation using real-world measurement data from the Massif Central tunnel in France further confirms that the model maintains excellent performance under practical, data-scarce, and long-range conditions, highlighting its strong potential for real-world deployment.

\section{Methodology}\label{sec:method}
\begin{figure*}[!t]
\vspace*{-1.0cm}
\centering

\subfloat[Model overview]{%
\includegraphics[width=0.48\textwidth]{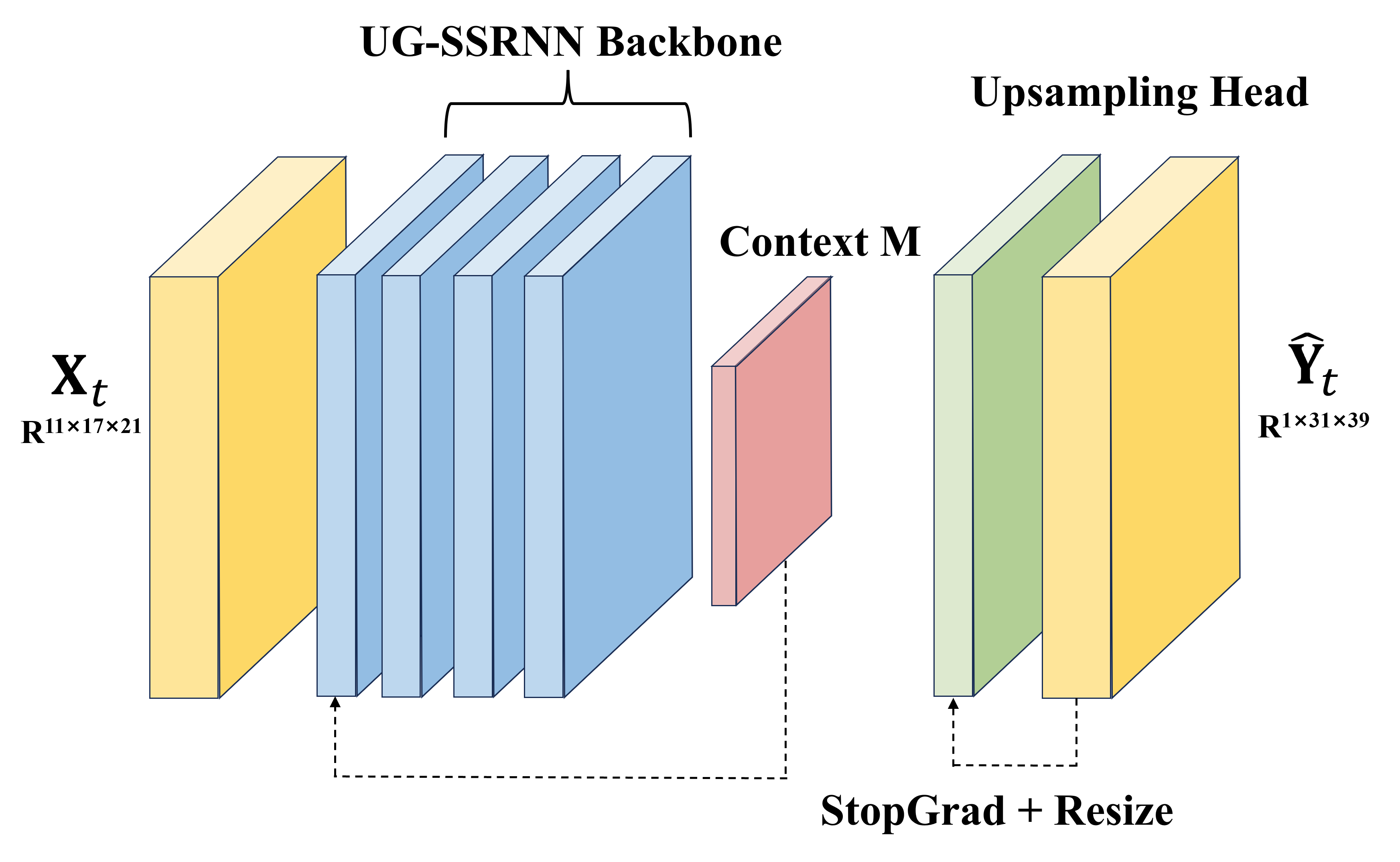}
}\hfill
\subfloat[Detailed module structure]{%
\includegraphics[width=0.48\textwidth]{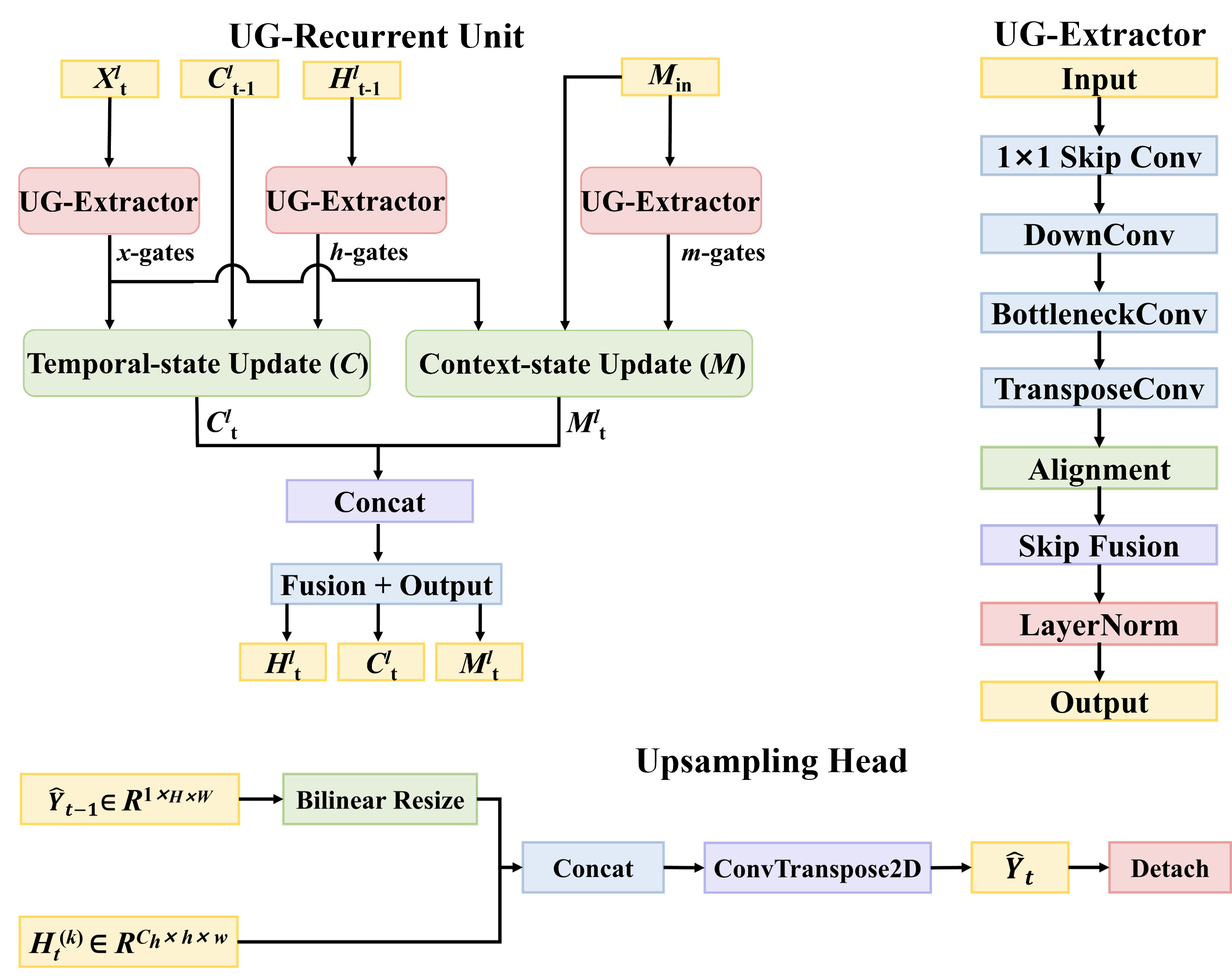}
}

\caption{Architecture of the proposed model.}
\label{fig:model_arch}
\end{figure*}

UG-SSRNN consists of sliding-window context encoding, convolutional recurrent propagation with a shared propagation-context state \(M\), and prediction-aware upsampling. \rev{The sliding-window input provides local longitudinal context for missing-slice recovery, the shared propagation-context state \(M\) preserves accumulated propagation information across recurrent steps, and the prediction-aware upsampling head uses the previous prediction to improve slice-to-slice continuity.} The model overview and detailed modules are shown in Fig.~\ref{fig:model_arch}. \rev{Given the encoded coarse-field sequence \(\{\mathbf{X}_t\}_{t=1}^{A}\), UG-SSRNN predicts the corresponding fine-mesh electromagnetic field sequence \(\{\hat{\mathbf{Y}}_t\}_{t=1}^{A}\).}

\subsection{Context Encoding with Sliding Window and Availability Mask}

Let \(\mathbf{x}_t\) denote the coarse-grid field slice at longitudinal index \(t\), and let \(m_t\in\{0,1\}\) indicate whether the target slice is available. The indicator is spatially replicated as a mask plane \(\mathbf{A}_t\) with \(\mathbf{A}_t\equiv m_t\). For each target index, a length-\(L\) local window and the mask plane are stacked as
\begin{equation}
\mathbf{X}_t=\big[\mathbf{x}_t,\mathbf{x}_{t+1},\ldots,\mathbf{x}_{t+L-1},\mathbf{A}_t\big],
\label{eq:input_tensor}
\end{equation}
where missing slices are zero-filled. For a raw sequence length \(T\), the effective input sequence is \(\{\mathbf{X}_t\}_{t=1}^{A}\) with \(A=T-(L-1)\). The reconstruction target is \(\{\mathbf{Y}_t\}_{t=1}^{A}\), and the network outputs \(\{\hat{\mathbf{Y}}_t\}_{t=1}^{A}\).

\subsection{UG-SSRNN Backbone and Propagation-Context State \(M\)}

 The backbone contains \(K\) stacked convolutional recurrent layers. At step \(t\), layer \(k\) maintains a hidden feature map \(\mathbf{H}_t^{(k)}\) and an accumulator state \(\mathbf{C}_t^{(k)}\). A shared propagation-context state \(\mathbf{M}_t\) is refined across layers and fed diagonally from the previous step:
\begin{align}
\mathbf{M}_t^{(0)} &= \mathbf{M}_{t-1}^{(K)}, \nonumber\\
\big(\mathbf{H}_t^{(k)},\mathbf{C}_t^{(k)},\mathbf{M}_t^{(k)}\big)
&=
\Phi^{(k)}\!\left(
\mathbf{H}_t^{(k-1)},\mathbf{H}_{t-1}^{(k)},
\mathbf{C}_{t-1}^{(k)},\mathbf{M}_t^{(k-1)}
\right).
\label{eq:ssrnn_update}
\end{align}
where \(\mathbf{H}_t^{(0)}=\mathbf{X}_t\), and \(\Phi^{(k)}(\cdot)\) denotes a convolutional gated update implemented with a lightweight UG-extractor. This design enables long-range propagation information to persist without increasing the input window length, while also capturing boundary-induced and multi-scale spatial patterns. After processing all layers, \(\mathbf{M}_t=\mathbf{M}_t^{(K)}\), and the top-layer hidden state \(\mathbf{H}_t^{(K)}\) is used for reconstruction.

\subsection{Prediction-Aware Upsampling Head}

At each step \(t\), the fine-grid field is reconstructed from the top-layer hidden state and the previous prediction:
\begin{equation}
\hat{\mathbf{Y}}_t=
\mathcal{G}\!\left(
\mathrm{Concat}\!\left[
\mathbf{H}_t^{(K)},
\mathrm{Down}\!\big(\mathrm{StopGrad}(\hat{\mathbf{Y}}_{t-1})\big)
\right]\right),
\label{eq:head_out}
\end{equation}
where \(\hat{\mathbf{Y}}_0=\mathbf{0}\), and \(\mathcal{G}(\cdot)\) is a lightweight transposed-convolution upsampling module. The feedback term improves slice-to-slice consistency during sequential reconstruction.

\subsection{Loss Function}

All fields are normalized using a global Z-score computed over the training set. The network is optimized by a time-weighted MAE:
\begin{equation}
\mathcal{L}=\frac{1}{B}\sum_{b=1}^{B}
\frac{\sum_{t=1}^{A}\omega_t\left\lVert \hat{\mathbf{Y}}^{(b)}_{t}-\mathbf{Y}^{(b)}_{t}\right\rVert_1}
{\sum_{t=1}^{A}\omega_t+\epsilon},
\label{eq:loss}
\end{equation}
where \(\omega_t\) upweights indices with available coarse observations.


\section{Numerical Examples}\label{Section III}

\subsection{Simulation and Dataset Generation}

\begin{figure}[t]
\centering

\subfloat[]{%
\includegraphics[width=0.23\textwidth]{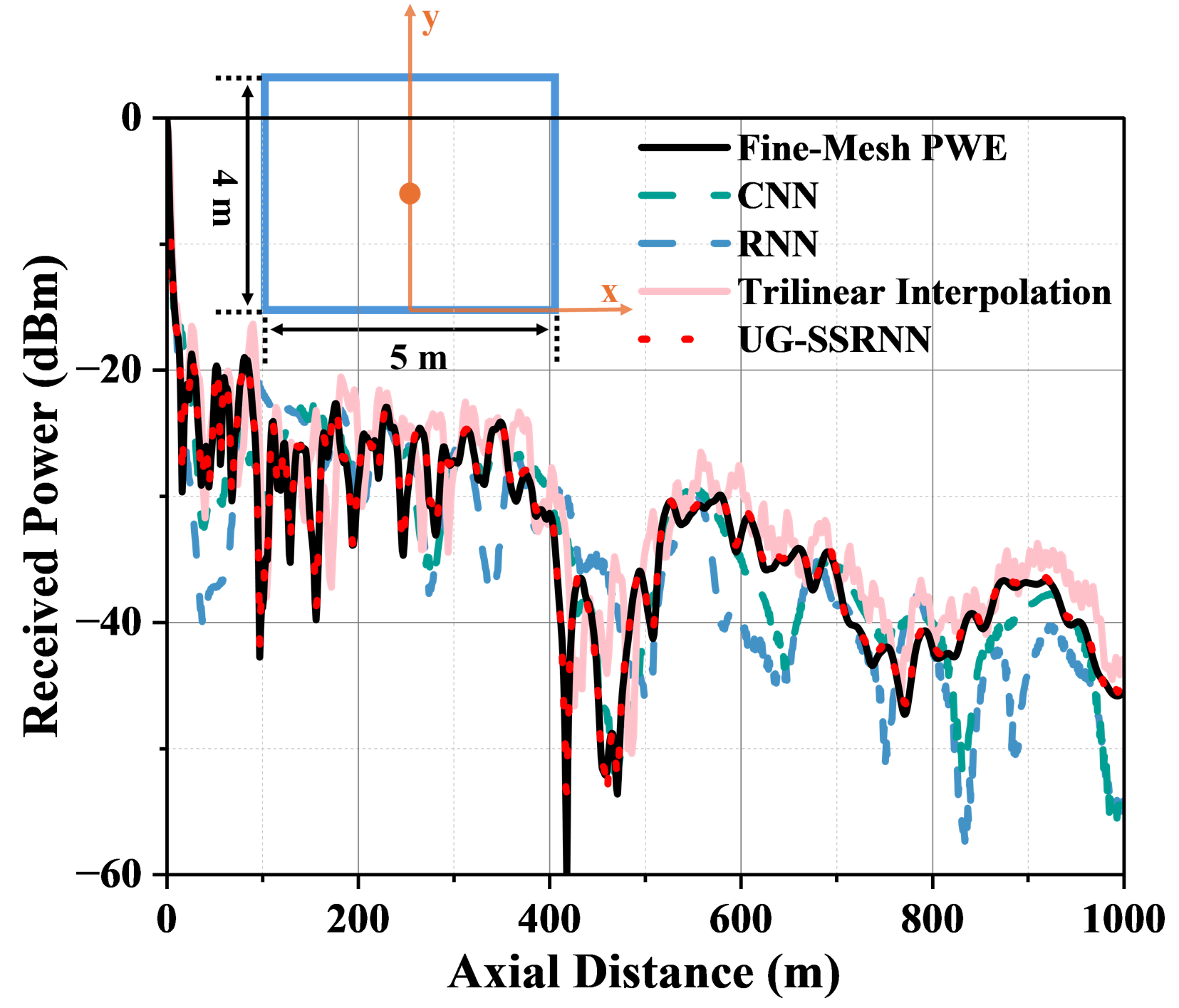}
}\hfill
\subfloat[]{%
\includegraphics[width=0.23\textwidth]{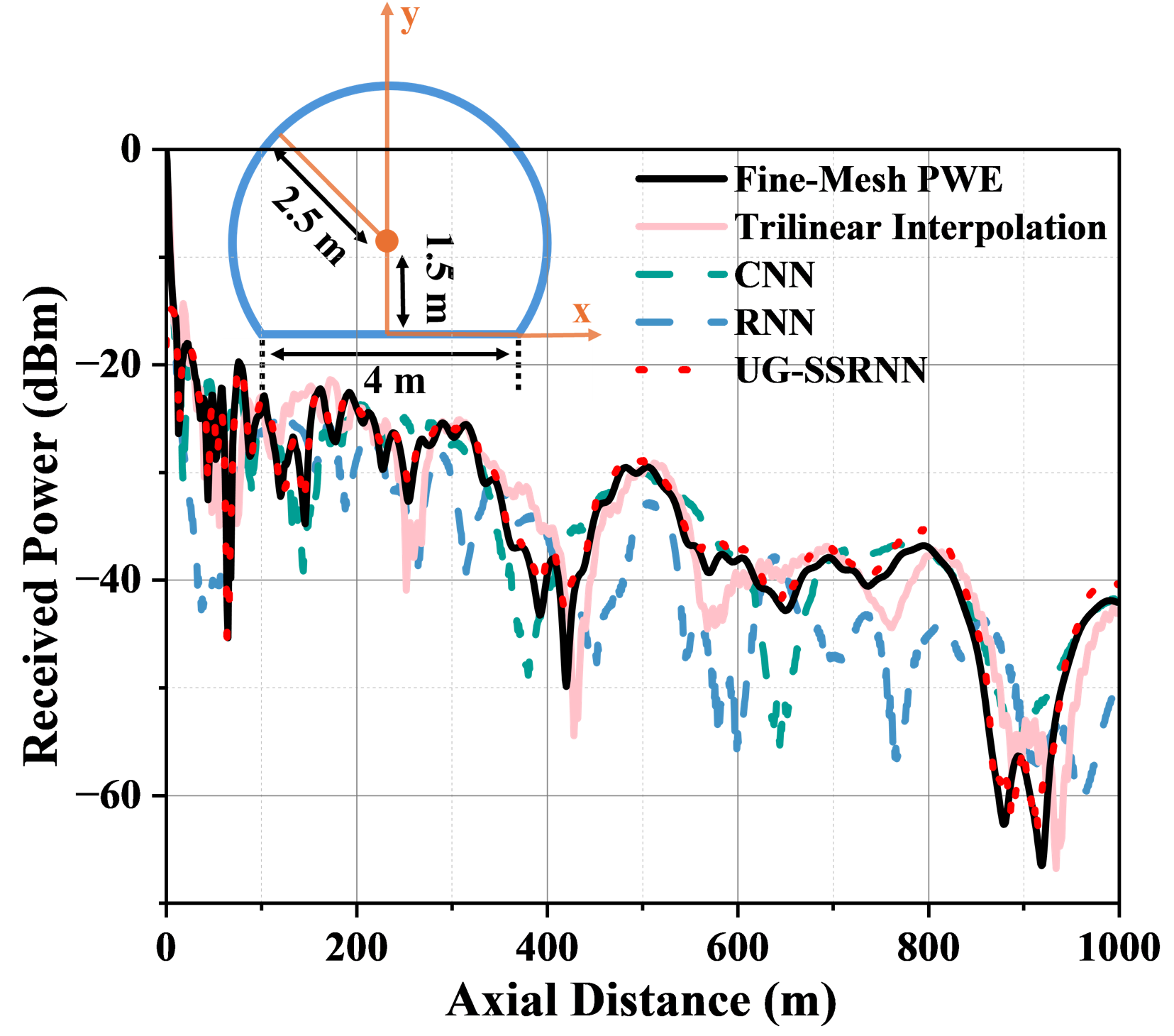}
}\hfill
\subfloat[]{%
\includegraphics[width=0.23\textwidth]{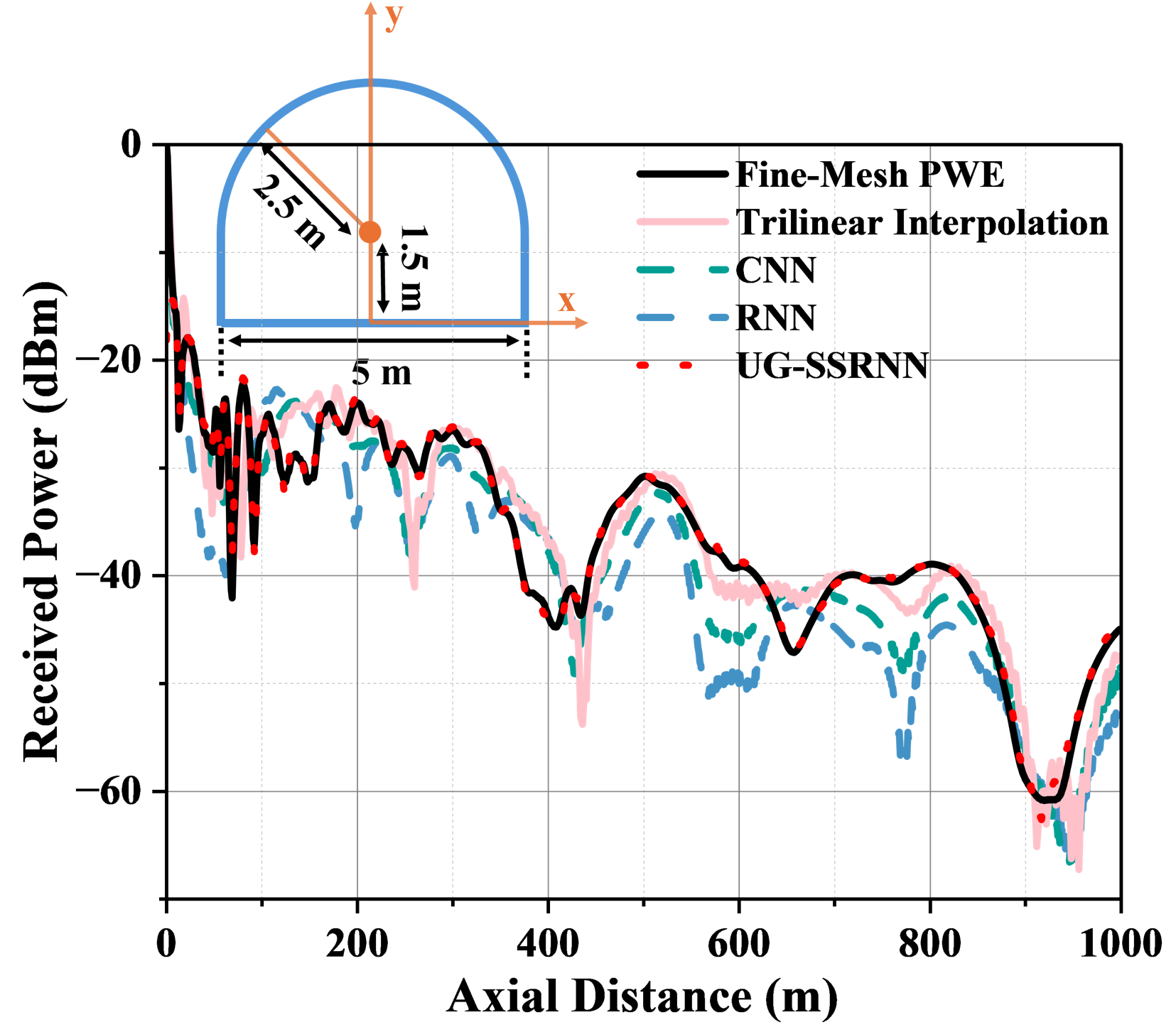}
}\hfill
\subfloat[]{%
\includegraphics[width=0.23\textwidth]{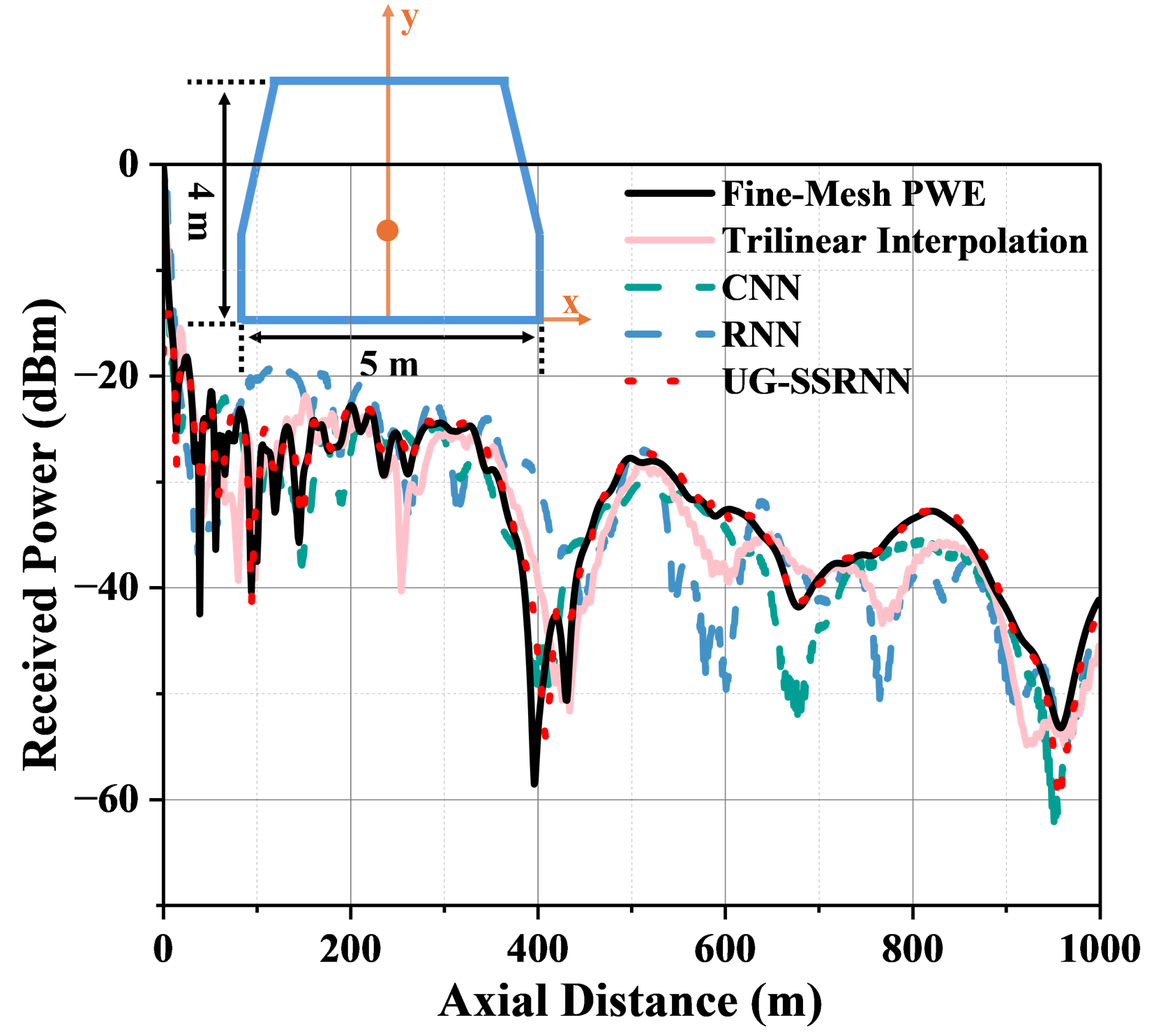}
}

\caption{\rev{Received-power comparison at 0.9~GHz for four tunnel geometries: (a) rectangular, (b) arched, (c) arched with vertical side walls, and (d) trapezoidal. Average MAE values for trilinear interpolation, 2-D CNN, RNN, and UG-SSRNN are 4.562, 2.887, 5.573, and 0.749~dB, respectively. Insets show the corresponding cross sections.}}

\label{fig:comparison}
\end{figure}

\rev{To evaluate the proposed model, paired coarse- and fine-mesh fields are generated using a parabolic wave equation (PWE)-based solver in a 1000 m tunnel. This choice is consistent with tunnel propagation physics, since parabolic equation/vector parabolic equation (PE/VPE) methods are efficient for long guiding structures and have been validated in railway tunnels, whereas full-wave finite-difference time-domain (FDTD) solvers are impractical for electrically large domains and ray tracing (RT) requires high-order reflections for long-distance convergence~\cite{7420622,popov19993d}.} \rev{As shown in the insets of Fig.~\ref{fig:comparison}, four representative tunnel cross sections are considered, including rectangular, arched, arched with vertical side walls, and trapezoidal geometries.} At the tunnel entrance, the electromagnetic field is excited by a transmitting antenna located at $(x_{\mathrm{TX}}, y_{\mathrm{TX}})$, modeled as a Gaussian beam with unit amplitude and a beam width corresponding to a standard deviation of $3\lambda$.

\begin{figure}[!t]
\centering
    \includegraphics[width=0.80\linewidth]{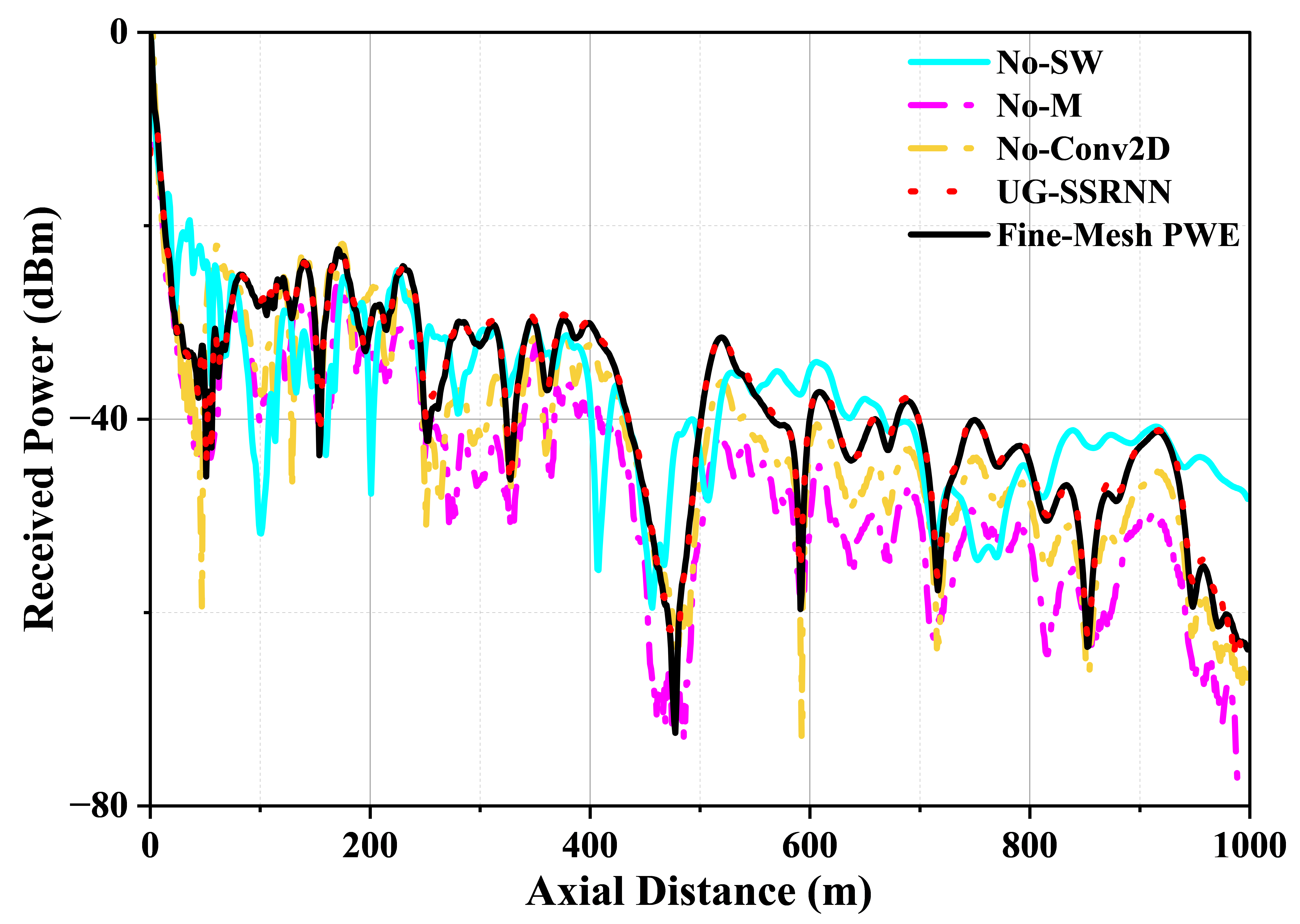}
    \caption{\rev{Received-power profiles in the rectangular tunnel for UG-SSRNN and ablated variants.}}
    \label{fig:ablation}
\end{figure}

The numerical simulations employ a multi-resolution discretization strategy. The coarse mesh adopts $\Delta x=\Delta y=0.8\lambda$ and $\Delta z=4.0\lambda$, while the fine reference mesh uses $\Delta x=\Delta y=0.4\lambda$ and $\Delta z=2.0\lambda$. \rev{The dataset is constructed by sampling tunnel shape, wall relative permittivity $\epsilon_r$, wall conductivity $\sigma$, and transmitter coordinates $(x_{\mathrm{TX}},y_{\mathrm{TX}})$. Unless otherwise specified, simulations are generated at 0.9~GHz with four tunnel shapes, $\epsilon_r\in\{5,7.5\}$, $\sigma\in\{0.01,0.1\}$~S/m, $x_{\mathrm{TX}}=0.5{:}0.5{:}2.0$~m, and $y_{\mathrm{TX}}=1.5{:}0.5{:}3.0$~m. The samples are split into 80\% for training and 20\% for testing, with part of the training set used for validation and model selection. Each input contains ten sliding-window coarse slices and one availability-mask channel, and the output is the corresponding fine-mesh slice. Hyperparameters are selected according to validation-set MAE and computational cost~\cite{bergstra2012random}. The final model uses $L=10$, four recurrent layers, 16 hidden channels, AdamW with learning rate $10^{-3}$, and batch size 1. It has 1.29M parameters and 0.138~GFLOPs per slice, and runs on an Intel Core i7-13700 CPU and an NVIDIA RTX 4090 GPU with 24~GB VRAM.}

\subsection{Ablation Study}

\rev{Fig.~\ref{fig:ablation} compares the full UG-SSRNN with three ablated variants in the rectangular tunnel: no sliding window (No-SW), no propagation-context state \(M\) (No-M), and no convolutional spatial modeling (No-Conv2D). No-SW shows the largest deviation from the fine-mesh PWE reference, confirming the importance of neighboring coarse slices for recovering missing longitudinal information. No-M weakens long-range consistency and fading-notch tracking, while No-Conv2D degrades local transverse reconstruction. The full UG-SSRNN provides the closest agreement with the reference profile, demonstrating the complementary roles of sliding-window context, propagation memory, and convolutional spatial modeling.}

\begin{figure}[ht]
	\vspace{-1cm}
	\centering
	\subfloat[UG-SSRNN]{\includegraphics[width=0.24\textwidth]{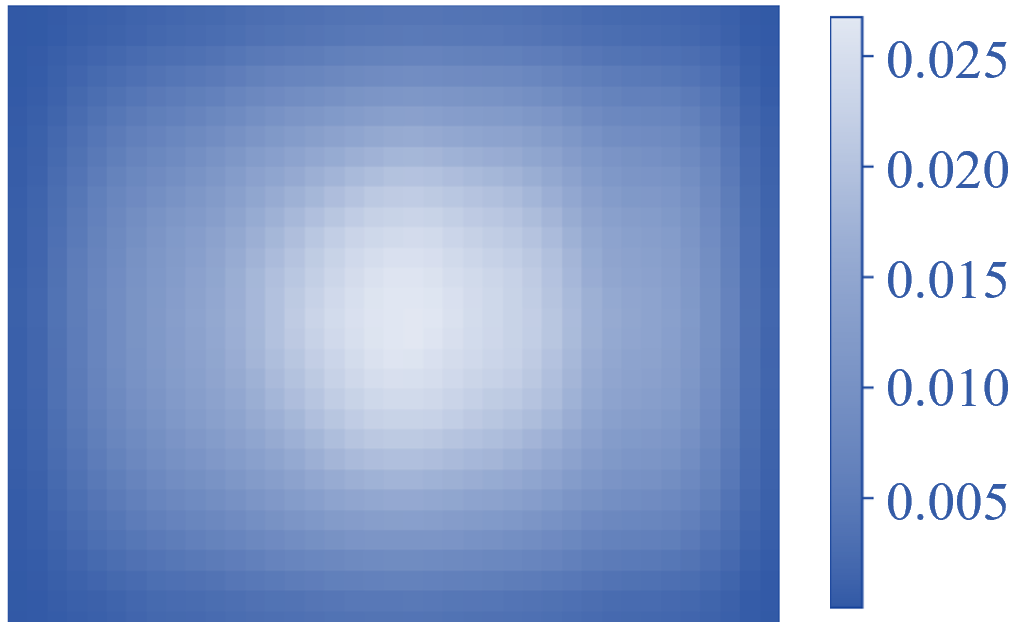}\label{fig:input500}} \hfill
	\subfloat[Fine-mesh PWE]{\includegraphics[width=0.24\textwidth]{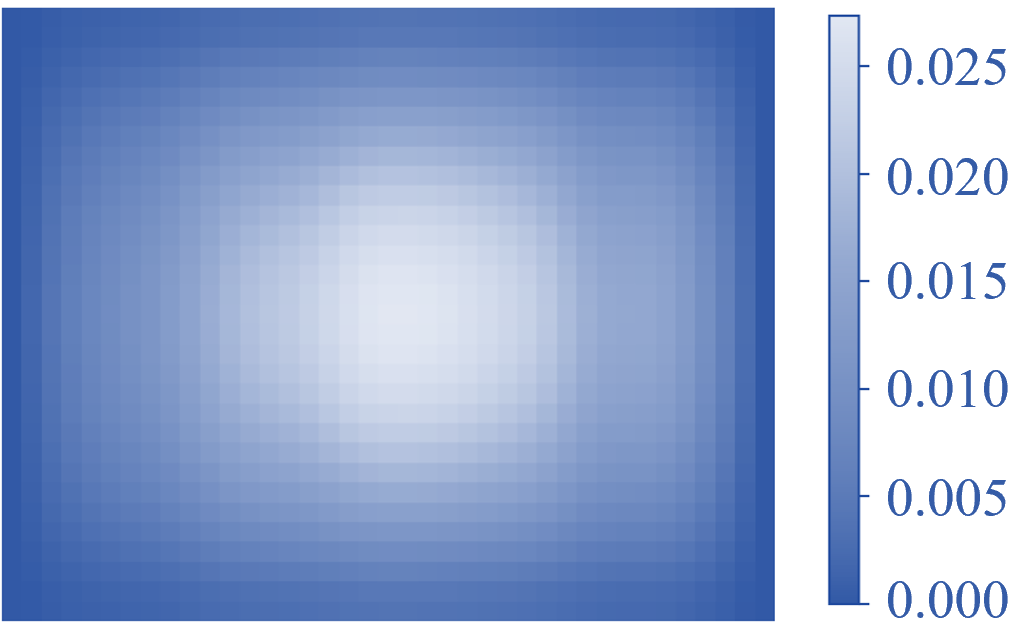}\label{fig:prediction500_48}} \hfill

	\subfloat[UG-SSRNN]{\includegraphics[width=0.24\textwidth]{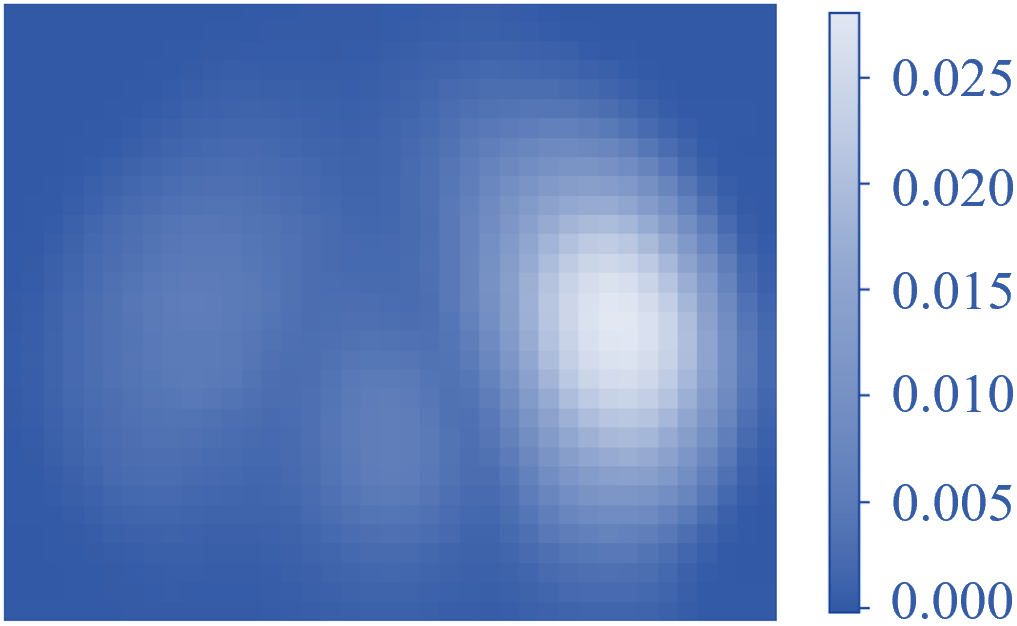}\label{fig:input1000_48}} \hfill
	\subfloat[Fine-mesh PWE]{\includegraphics[width=0.24\textwidth]{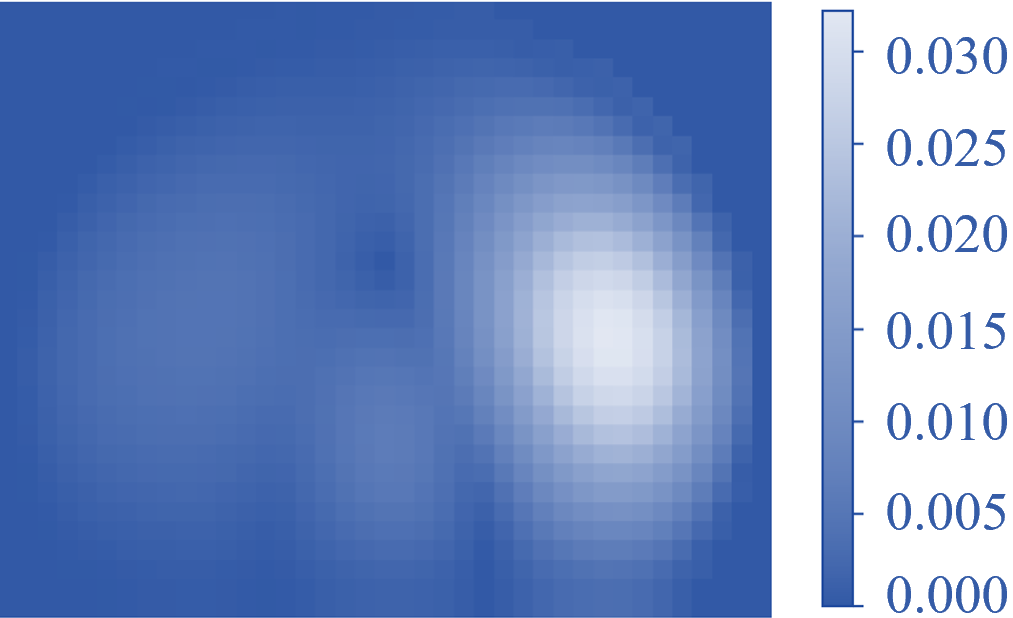}\label{fig:predict00_458}} \hfill

    \subfloat[UG-SSRNN]{\includegraphics[width=0.24\textwidth]{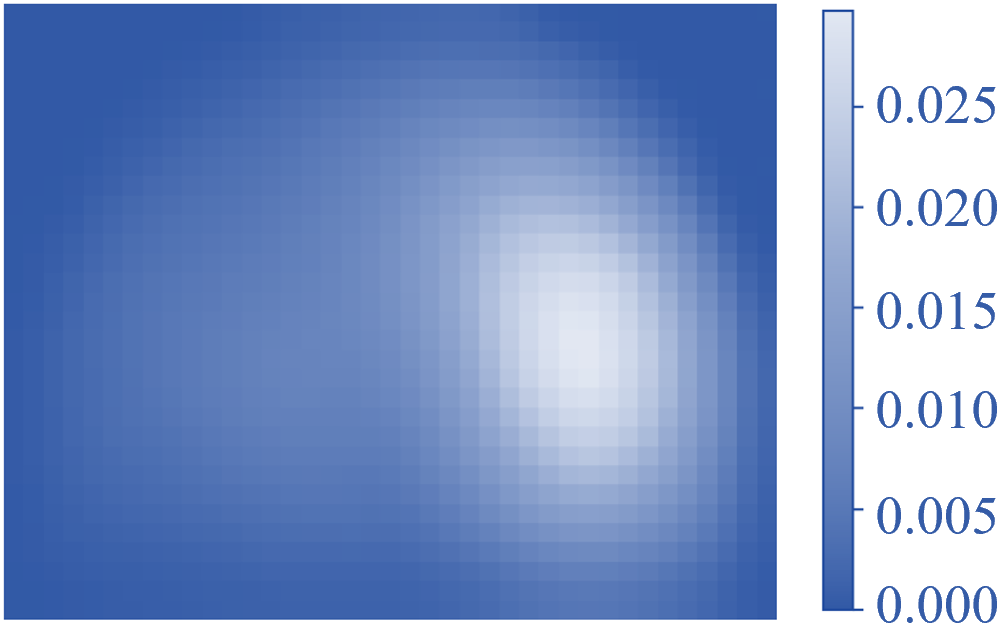}}\label{fig:500_458} \hfill
	\subfloat[Fine-mesh PWE]{\includegraphics[width=0.24\textwidth]{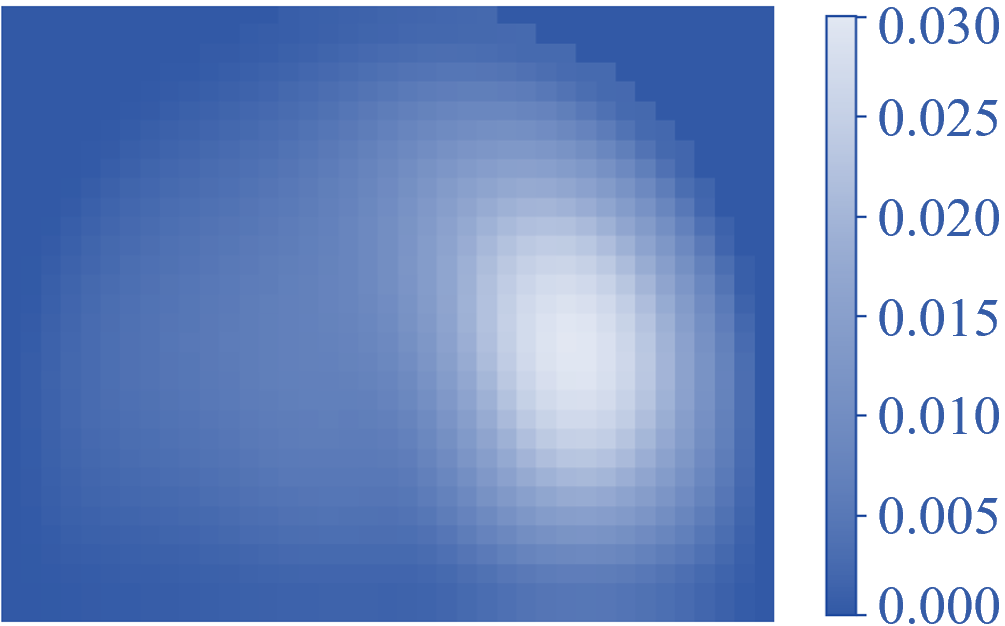}\label{fig:pred00_458}} \hfill

	\subfloat[UG-SSRNN]{\includegraphics[width=0.24\textwidth]{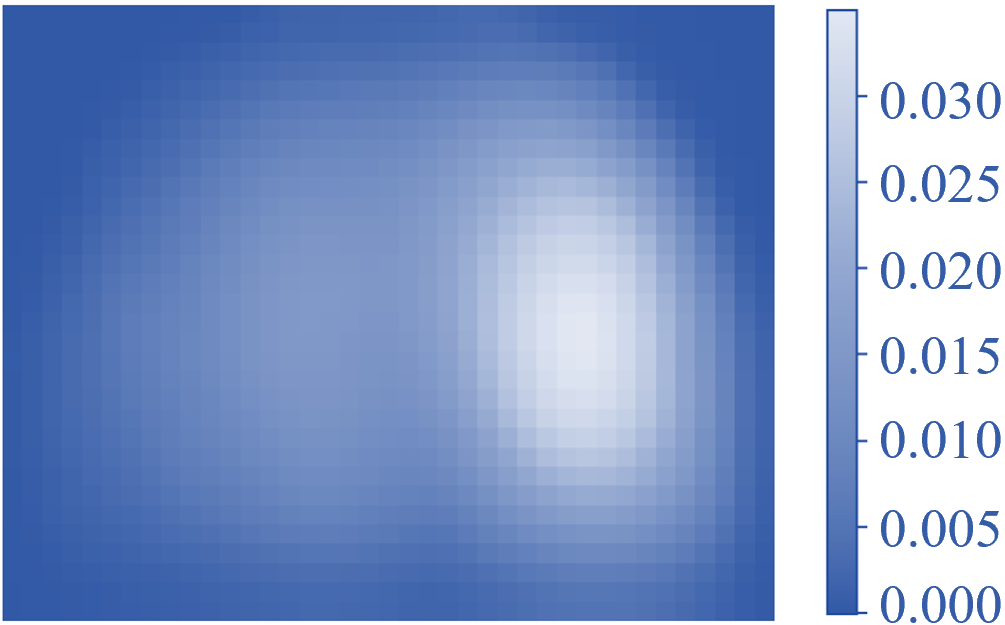}}\label{finput1000_458} \hfill
	\subfloat[Fine-mesh PWE]{\includegraphics[width=0.24\textwidth]{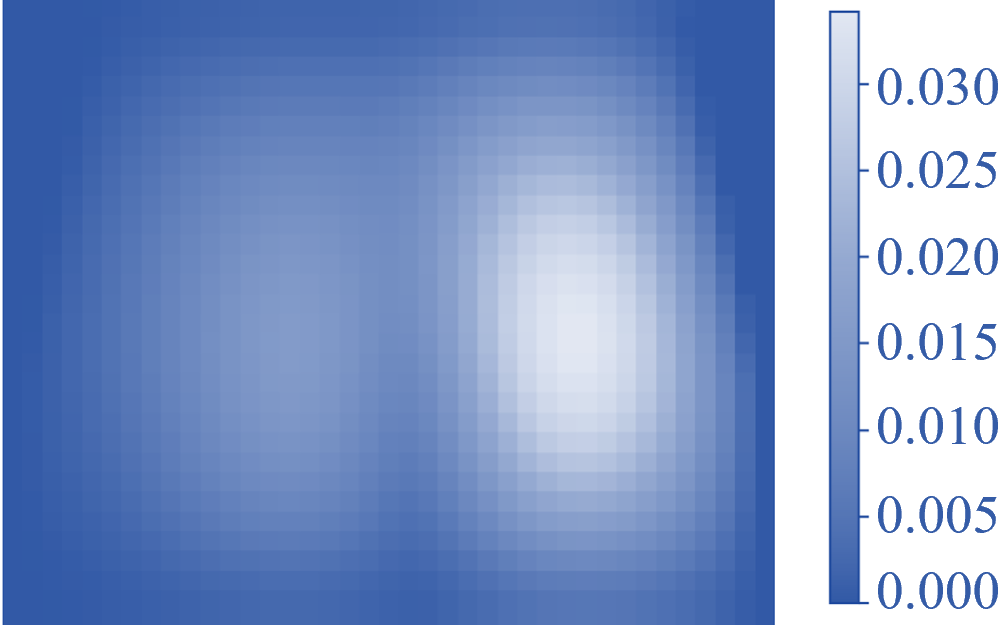}\label{fiction1058}} \hfill

\caption{Comparison of cross-sectional distributions at 500 m between fine-mesh PWE results and the proposed model predictions for four tunnel geometries: (a)–(b) rectangular, (c)–(d) arched, (e)–(f) arched with vertical side walls, and (g)–(h) trapezoidal.}
\label{fig:two_rows_four_cols}
\end{figure}

\subsection{Comparison Experiments}

\rev{We compare UG-SSRNN with 3-D trilinear interpolation, a 2-D convolutional neural network (CNN), and a recurrent neural network (RNN) under the same train-test split and evaluation range, using the fine-mesh PWE result as the reference. As shown in Fig.~\ref{fig:comparison}, trilinear interpolation mainly performs fixed resampling of the coarse-mesh field and misses several local fading notches. Its deviation becomes more evident in the far-distance region because interpolation cannot recover fine-scale fading details that are absent or shifted in the coarse-mesh PWE field. The CNN and RNN baselines reduce part of the error but still show visible deviations from the reference profiles. In contrast, UG-SSRNN more closely follows both the overall attenuation trend and the local fading variations across the four tunnel geometries. Fig.~\ref{fig:two_rows_four_cols} further confirms that the proposed model preserves the transverse field distribution at 500~m, demonstrating its joint longitudinal and transverse reconstruction capability.}

\begin{figure}[!t]
    \vspace{-0.8cm}
    \centering
    \subfloat[]{%
    \includegraphics[width=0.48\linewidth]{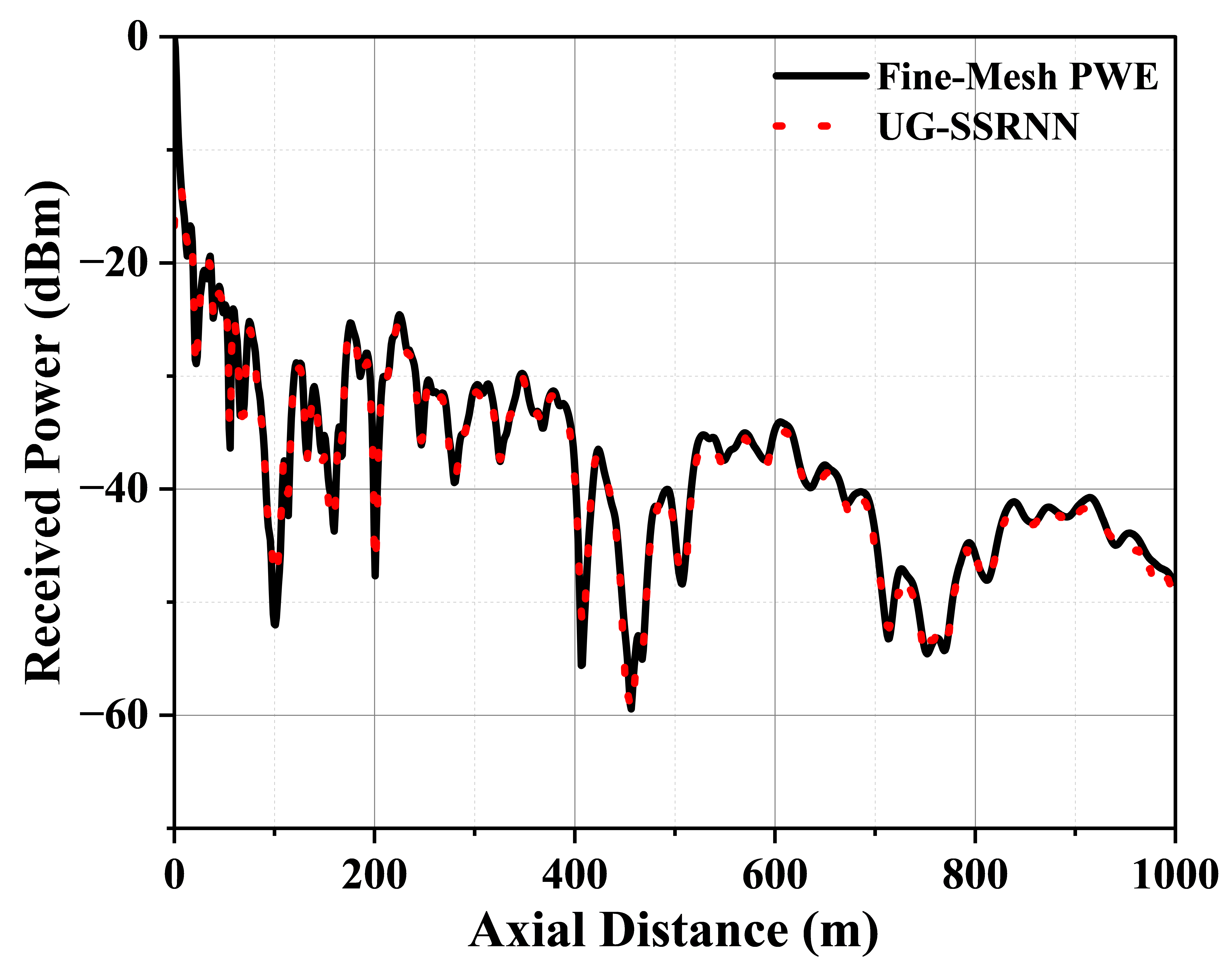}
    \label{fig:gen_material}
    }\hfill
    \subfloat[]{%
    \includegraphics[width=0.48\linewidth]{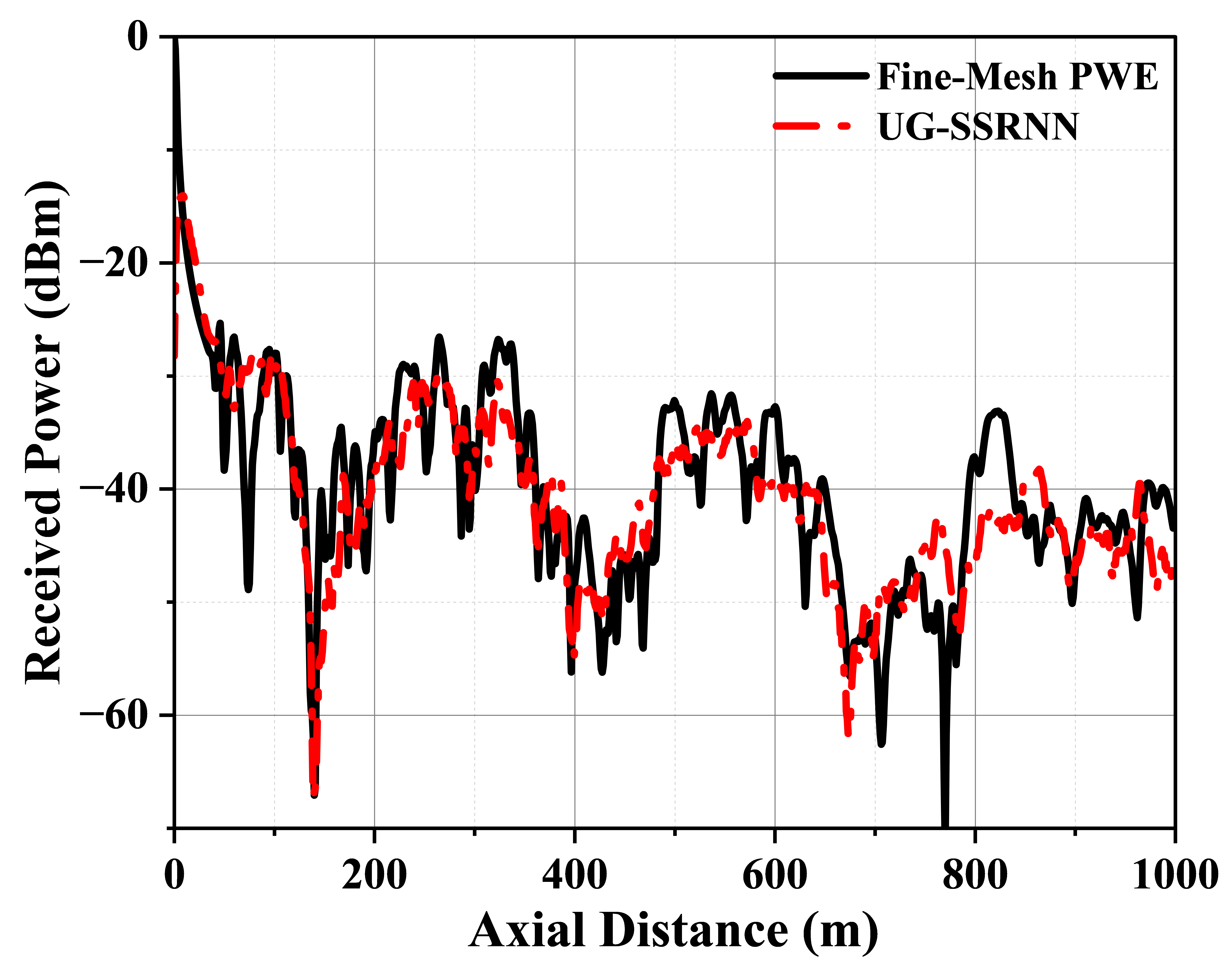}
    \label{fig:gen_frequency}
    }
    \vspace{-0.2cm}
    \caption{\rev{Generalization experiments of UG-SSRNN: (a) unseen-material test and (b) unseen-frequency test at 2.4~GHz.}}
    \label{fig:generalization}
    \vspace{-0.2cm}
\end{figure}

\subsection{Generalization Experiments}
\rev{To further evaluate cross-condition generalization, we conduct unseen-material and unseen-frequency tests, as shown in Fig.~\ref{fig:generalization}. In the unseen-material test, the material-parameter combination $(\epsilon_r = 10,\sigma = 0.001)$ is completely excluded from the training and validation sets and used only for testing. UG-SSRNN achieves MAE, MAPE, and RMSE of 0.302~dB, 1.874\%, and 0.657~dB, respectively. In the cross-frequency test, the model is trained at 0.9~GHz and directly tested at 2.4~GHz, achieving MAE, MAPE, and RMSE of 3.964~dB, 9.990\%, and 5.048~dB, respectively. These results indicate that the learned coarse-to-fine reconstruction remains accurate under unseen material parameters and retains reasonable robustness under frequency variation. This robustness is attributed to the sliding-window input and propagation-context state \(M\), which encode longitudinal attenuation and fading evolution, together with convolutional spatial modeling, which captures transverse boundary-induced field patterns.}

\section{Application: Massif Central Tunnel}\label{Section IV}

\begin{figure}[!t]
    \vspace{-0.6cm}
    \centering
    \includegraphics[width=0.8\linewidth]{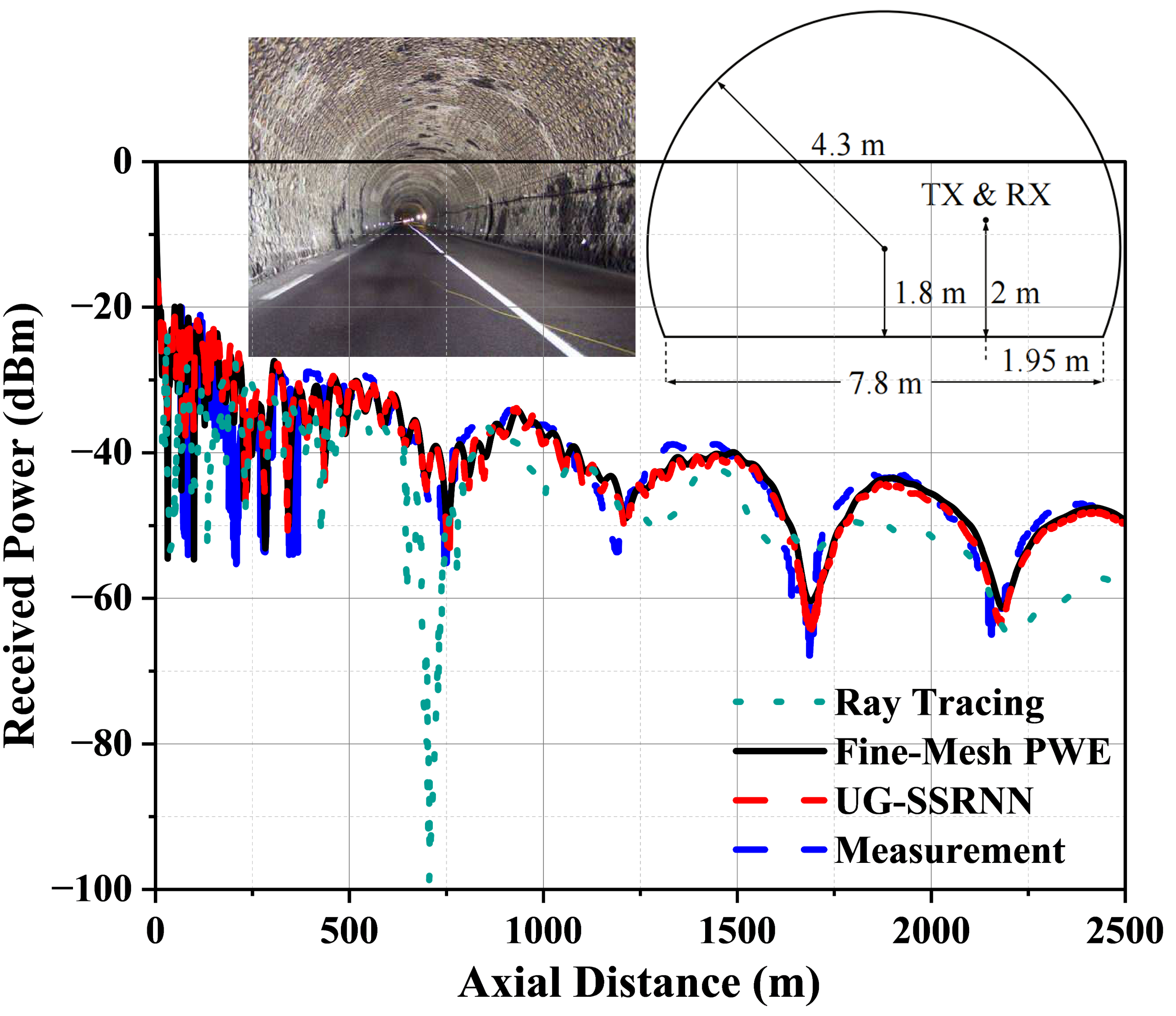}
    \caption{\rev{Received power predictions along the Massif Central tunnel, including the converged RT result with 25 reflections, fine-mesh PWE reference, UG-SSRNN prediction, and measurement data. Insets show the on-site photograph and the cross-sectional geometry of the tunnel.}}
    \label{fig:ugstrnn}
\end{figure}

The proposed model is validated in the Massif Central tunnel, a large-scale enclosed environment with strong boundary effects and long-range dependencies. Unlike the synthetic setups in Section III, the field evolution is smoother yet structured due to the tunnel geometry and material composition.
The tunnel is modeled as a long enclosed structure primarily composed of stone and reinforced concrete, as shown in Fig.~\ref{fig:ugstrnn}, with boundary parameters $\varepsilon_r = 5$ and $\sigma = 0.01~\mathrm{S/m}$. The carrier frequency is set to 0.9~GHz. Coarse-mesh discretization uses $\Delta x=\Delta y=0.2\lambda$ and $\Delta z=0.5\lambda$, while fine-mesh discretization uses $\Delta x=\Delta y=0.1\lambda$ and $\Delta z=0.25\lambda$.

Fig.~\ref{fig:ugstrnn} compares the predicted received-power profile with the fine-mesh PWE reference, a converged Ray Tracing (RT) result with 25 reflections, and the available measurement data. \rev{RT captures the overall attenuation trend but shows stronger local deviations in several fading regions, whereas UG-SSRNN closely follows the fine-mesh PWE reference and remains consistent with the measured profile.} The MAE, MAPE, and RMSE values of UG-SSRNN with respect to the fine-mesh PWE reference are 1.343~dB, 3.794\%, and 2.201~dB, respectively. \rev{The network inference time is 9.27~s, and the coarse-mesh PWE input generation requires 76.55~s. In comparison, the conventional fine-mesh PWE simulation requires 879.70~s.}

\section{Conclusion}
\rev{This letter presented UG-SSRNN for 3-D coarse-to-fine reconstruction of tunnel electromagnetic fields. It combines sliding-window context encoding, convolutional recurrent propagation with a shared propagation-context state, and prediction-aware upsampling to capture both longitudinal evolution and transverse spatial details. Experiments show that UG-SSRNN outperforms 3-D trilinear interpolation, 2-D CNN, and RNN baselines, while ablation confirms the contributions of temporal context, propagation memory, and spatial modeling. The unseen-material, unseen-frequency, and Massif Central tunnel tests further demonstrate robustness under material, frequency, and practical tunnel conditions. Compared with fine-mesh PWE, the proposed method achieves accurate reconstruction with much lower computational cost.}

\label{sec:conclusion}

\bibliographystyle{IEEEtran}
\bibliography{MyBibliography}





\end{document}